\documentclass[
    superscriptaddress,
    reprint,
    showkeys,
    showpacs,
    amsmath,
    amssymb,
    aps,
    prb,
    twocolumn,
    floatfix
]{revtex4-1}
\usepackage{amsmath}
\usepackage{physics}
\usepackage{siunitx}
\usepackage{graphicx}
\usepackage{dcolumn}
\usepackage{bm}
\usepackage{braket}
\usepackage[caption=false]{subfig}
\usepackage{float}
\usepackage{booktabs}
\usepackage{multirow}
\usepackage{tabularx}
\usepackage{color}
\definecolor{link}{RGB}{57,106,177}
\definecolor{blue}{RGB}{0,0,0}
\newcommand{\ped}[1]{\ensuremath{_{\rm #1}}}
\newcommand{\apex}[1]{\ensuremath{^{\rm #1}}}

\newcommand{\mywidth}{1.0\columnwidth}
\newcommand{\mysmallwidth}{0.8\columnwidth}
\usepackage{hyperref}
\hypersetup{
    colorlinks=true,
    linkcolor=link,
    citecolor=link,
    urlcolor=link
}
\AtBeginDocument{
\heavyrulewidth=.08em
\lightrulewidth=.05em
\cmidrulewidth=.03em
\belowrulesep=.65ex
\belowbottomsep=0pt
\aboverulesep=.4ex
\abovetopsep=0pt
\cmidrulesep=\doublerulesep
\cmidrulekern=.5em
\defaultaddspace=.5em
}
\begin{document}
\title{Migdal-Eliashberg theory of multi-band high-temperature superconductivity in field-effect-doped hydrogenated (111) diamond}
\author{Davide Romanin}
\email{davide.romanin@polito.it}
\affiliation{
    Department of Applied Science and Technology, Politecnico di Torino, 10129 Torino, Italy
}
\author{Giovanni A. Ummarino}
\affiliation{
    Department of Applied Science and Technology, Politecnico di Torino, 10129 Torino, Italy
}
\affiliation{National Research Nuclear University MEPhI (Moscow Engineering Physics Institute), Kashira Hwy 31, Moskva 115409, Russia
}
\author{Erik Piatti}
\email{erik.piatti@polito.it}
\affiliation{
    Department of Applied Science and Technology, Politecnico di Torino, 10129 Torino, Italy
}

\begin{abstract}
We perform single- and multi-band Migdal-Eliashberg (ME) calculations with parameters exctracted from density functional theory (DFT) simulations to study superconductivity in the electric-field-induced 2-dimensional hole gas at the hydrogenated (111) diamond surface. We show that according to the Eliashberg theory it is possible to induce a high-T$_{\text{c}}$ superconducting phase when the system is field-effect doped to a surface hole concentration of $6\times10^{14}\,$cm$^{-2}$, where the Fermi level crosses three valence bands. Starting from the band-resolved electron-phonon spectral functions $\alpha^2F_{jj'}(\omega)$ computed ab initio, we iteratively solve the self-consistent isotropic Migdal-Eliashberg equations, in both the single-band and the multi-band formulations, in the approximation of a constant density of states at the Fermi level. In the single-band formulation, we find T$_{\text{c}}\approx40$~K, which is enhanced between $4\%$ and $8\%$ when the multi-band nature of the system is taken into account. We also compute the multi-band-sensistive quasiparticle density of states to act as a guideline for future experimental works. 
\end{abstract}


\maketitle

\section{\label{sec:intro}Introduction}

Superconductivity (SC) in diamond was first reported in 2004 by E. A. Ekimov et al.~\cite{EkimovNature2004} by boron (B) substitution of carbon (C) atoms, leading to a hole-type conductivity and to a transition temperature $T\ped{c}=4$~K~\cite{EkimovNature2004, BustarretPSSA2008} at a dopant concentration $n\ped{B}\approx4-5\cdot10^{21}$~cm$^{-3}$. It was then confirmed theoretically that the B-doped bulk diamond is a conventional phonon-mediated superconductor, and that by increasing the B concentration it would be possible to enhance $T\ped{c}$~\cite{BlasePRL2004, BoeriJPCS2006, GiustinoPRL2007, MoussaPRB2008, CalandraPRL2008, CarusoPRL2017}. From the experimental point of view, however, the solubility limit~\cite{ChenAPL1999, SolozhenkoPRL2009} of B in the closely-packed diamond structure prevents increasing the doping further, thus limiting the highest achievable $T\ped{c}$. An alternative route {\color{blue}to induce hole-type conductivity in diamond is by hydrogen (H) termination followed by exposure to electron-accepting molecules such as air moisture~\cite{LandstrassAPL1989, MaierPRL2000, StrobelNature2004}, leading to the formation of a surface-bound high-mobility two-dimensional hole gas whose properties have been extensively investigated~\cite{LandstrassAPL1989, MaierPRL2000, StrobelNature2004, KawaradaSSR1996, NebelDRM2002, NebelDMR2004, EdmondsPRB2010, EdmondsNL2015}.} Furthermore, the carrier density in hydrogenated diamond can be increased by means of electric field-effect doping: The configuration is that of a field-effect transistor (FET), i.e. an asymmetric capacitor where the material under study is separated from a metallic gate by a suitable dielectric. The application of a finite voltage between the gate and the sample ($V\ped{G}$) allows to accumulate charges-carriers in the first few layers of the sample (whose sign depends on that of $V\ped{G}$) which screen the resulting applied electric field [Fig.~\ref{fig:el_phon_DFT}(a)]. While a solid dielectric allows for an effective doping of the order of $\sim10^{12}-10^{13}$~cm$^{-2}$, the use of an ionic liquid or a polymer-electrolyte solution can increase it to $\sim10^{14}-10^{15}$~cm$^{-2}$~\cite{UenoJPSJ2014, DagheroPRL2012, PiattiJSNM2016, LiNature2016, PiattiPRB2017, XiPRL2016, Gonnelli2DMater2017, PiattiAPL2017, CostanzoNatNano2018, PiattiApSuSc2018mos2, PiattiPRM2019}. In the specific case of diamond, electrolyte-based gating can routinely tune hole dopings up to $\sim 7\times10^{13}$~cm\apex{-2}~\cite{PiattiLTP2019, YamaguchiJPSJ2013, TakahidePRB2014, AkhgarNL2016, TakahidePRB2016, AkhgarPRB2019, PiattiApSuSc2020}, and a proper engineering of the surface properties can extend the range at least to few units in $10^{14}$~cm\apex{-2}~\cite{PiattiLTP2019, PiattiApSuSc2020, PiattiEPJ2019}.

Previous density functional theory (DFT) simulations~\cite{PiattiLTP2019, NakamuraPRB2013, SanoPRB2017, RomaninAPSUSC2019} showed that it is possible to induce a SC phase transition in hydrogenated diamond surfaces by hole-doping in the FET configuration. In these studies, the critical temperature $T\ped{c}$ was estimated using the McMillan/Allen-Dynes~\cite{McMillanPR1968, AllenPRB1975} formula. In particular, in Refs.~\onlinecite{NakamuraPRB2013, SanoPRB2017} the authors studied the hydrogenated diamond (110) surface, and predicted a SC phase transition at $T\ped{c}\sim1$~K. In Ref.~\onlinecite{RomaninAPSUSC2019}, on the other hand, it was shown that the hydrogenated diamond (111) surface develops multi-band SC with a much higher $T\ped{c}\sim30$~K. However, the McMillan/Allen-Dynes formula only gives qualitative results: A more accurate estimate of $T\ped{c}$ necessitates solving the Migdal-Eliashberg (ME) equations instead~\cite{EliashbergSovPhys1960, EliashbergZhTeor1960}.

In this work -- starting from the results of Ref.~\onlinecite{RomaninAPSUSC2019} -- we study the electric-field-induced high-$T\ped{c}$ SC in the hydrogenated diamond (111) surface via the full self-consistent solution of the ME equations. In particular, we focus on the role of the multi-band Fermi surface on the SC pairing and $T\ped{c}$. In Sec.~\ref{sec:model} we summarize the main DFT results obtained by Ref.~\onlinecite{RomaninAPSUSC2019} at the field-induced hole doping of $n\ped{dop}=6\times10^{14}$ cm$^{-2}$, and introduce the multi-band Eliashberg equations on the imaginary axis, discussing also the Ansatz made for the effective electron-electron (e-e) interaction. In Sec.~\ref{sec:results} we move to the solution of the isotropic Eliashberg equations, first in the single-band approximation and then including the multi-band nature of the system (always in the approximation of a constant density of states at the Fermi level). We also determine the quasiparticle density of states from the analytic continuation of the SC gaps to the real axis.

\bigskip

\section{\label{sec:model}Model and Methods}

\subsection{DFT computation of the electron-phonon matrix elements}
\begin{figure}[]
\centering
     \includegraphics[width=\mywidth]{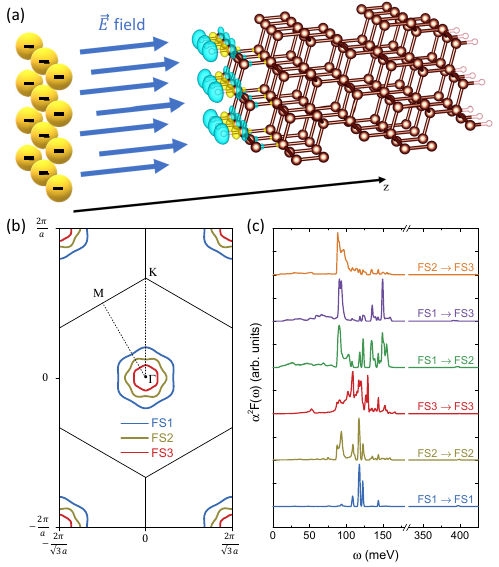}
      \caption
      {(a) Schematic view of the hydrogenated diamond (111) surface in the FET configuration. Brown spheres are C atoms, while pink spheres are H atoms. Light-blue (yellow) regions represents the induced hole (electron) charge densities on the surface. Blue arrows denote the electric field ($\vb{E}$) normal to the surface; (b) Fermi surface of the hydrogenated diamond (111) surface doped with $n\ped{dop}=6\times10^{14}$ cm$^{-2}$. FS1, FS2 and FS3 label the three different Fermi sheets; (c) Inter- and intra-band Eliashberg functions, normalized to 1, of the hydrogenated diamond (111) surface doped at $n\ped{dop}=6\times10^{14}$ cm$^{-2}$. The curves are in arbitrary units and vertically shifted for clarity.}
   \label{fig:el_phon_DFT}
\end{figure}

In Ref.~\onlinecite{RomaninAPSUSC2019} it was shown that a SC phase transition can be induced in the hydrogenated diamond (111) surface via field-effect doping at a hole concentration of $n\ped{dop}=6\times10^{14}$ cm$^{-2}$. This was done by performing DFT calculations in the proper field-effect geometry with Quantum ESPRESSO~\cite{QE, QE_2, SohierPRB2017} as described in Ref.~\onlinecite{SohierPRB2017} in the case of graphene: This is a very versatile approach, which has been employed to reliably calculate the properties of many different materials from first principles, including  zirconium nitride chloride~\cite{BrummePRB2014}, arsenene and phosphorene~\cite{SohierPRM2018}, transition-metal dichalcogenides~\cite{SohierPRM2018, BrummePRB2015, BrummePRB2016, SohierPRM2018, PiattiJPCM2019, RomaninJAP2020}, niobium nitride~\cite{PiattiApSuSc2018}, and silicon~\cite{RomaninNC2020}. For field-effect doped (111) diamond, all the computational details can be found in Ref.~\onlinecite{RomaninAPSUSC2019}. 
{\color{blue}As shown in Fig.~\ref{fig:el_phon_DFT}(a), we have considered a slab made of 14 C layers, terminated on both side by H atoms in order to saturate dangling bonds and avoid the surface reconstruction~\cite{PiattiLTP2019, RomaninAPSUSC2019}, for a total of 16 atoms in the unit cell. The presence of the H atoms on both side of the slab is also necessary to have a symmetric system as prescribed by the field-effect geometry implementation of Ref.~\cite{SohierPRB2017}, where the metallic gate is represented by a planar uniform distribution of point charges separated from the slab by a potential barrier to prevent charge spilling. Approximatively 30~\AA~of vacuum has also been inserted between different repeated images to avoid interferences between them due to the periodic boundary conditions.}
 
At $n\ped{dop}=6\times10^{14}$ cm$^{-2}$, the field-induced two-dimensional hole gas (2DHG) is multi-band in nature, since at the Fermi level (we set $E\ped{F}=0$) three holonic bands are crossed. As a consequence, the Fermi surface is composed of three non-crossing concentric hole pockets [Fig.~\ref{fig:el_phon_DFT}(b)] centered around $\vb{\Gamma}$ (i.e. the center of the Brillouin zone), which appear to be isotropic in $\vb{k}$-space (where $\vb{k}$-points denote the electron momenta). Tab.~\ref{tab:DOS} summarizes the total electronic density of states $N_{\sigma}(0)$ and the DOS for each band $N_{\sigma,j}(0)$ at the Fermi level, the corresponding band-averaged and band-resolved Fermi wavevectors, $\langle k\ped{F} \rangle$ and $\langle k\ped{F} \rangle_j$, and the band-averaged and band-resolved Fermi velocities, $\langle v\ped{F} \rangle$ and $\langle v\ped{F} \rangle_j$. Here, $j=1,2,3$ is the band index as labelled in Fig.~\ref{fig:el_phon_DFT}(b). We also stress that, since the field-induced 2DHG is strongly confined to the surface~\cite{RomaninAPSUSC2019}, the DOS is step-like: The assumption of constant DOS at the Fermi level is thus satisfied. The band-resolved Fermi velocities are calculated by averaging over each Fermi surface as:
\begin{equation}
\langle v\ped{F} \rangle_j = \frac{1}{2\pi}\int_0^{2\pi}d\theta\, \frac{1}{\hbar} \frac{\partial\epsilon_{j,\textbf{k}}}{\partial k}\biggr\rvert_{\epsilon_{j,\textbf{k}}=E\ped{F}}
\end{equation}
where $\hbar$ is the reduced Planck constant,  $\epsilon_{j,\textbf{k}}$ is the energy dispersion of the $j$-th band in $\textbf{k}$-space, $k=|\textbf{k}|$ and $\theta$ is the in-plane angular coordinate in $\textbf{k}$-space. The band-averaged Fermi velocity is the weighted average of the band-resolved Fermi velocities by the corresponding DOS at the Fermi level:
\begin{equation}
\langle v\ped{F} \rangle=\frac{\sum_{j}N_{\sigma,j}(0)\langle v\ped{F} \rangle_j}{\sum_{j}N_{\sigma,j}(0)}
\end{equation}
%
%
\begin{table}[b]
  \begin{tabular*}{\mywidth}{c @{\extracolsep{\fill}}ccc}
  \toprule
  Global	& $N_{\sigma}(0)$	& $\langle k\ped{F} \rangle$ & $\langle v\ped{F} \rangle$	\\
  \midrule
    	& 0.3936	& 3.62	& 3.58	\\
    \midrule
  Band-resolved    & $N_{\sigma,j}(0)$	& $\langle k\ped{F} \rangle_j$ & $\langle v\ped{F} \rangle_j$	\\
    \midrule
    Band 1	& 0.1783	& 4.81	& 3.71 \\
    Band 2	& 0.1088	& 3.26	& 4.28	\\
	Band 3	& 0.1059	& 2.02	& 2.65	\\
	\midrule
      & eV\apex{-1}~spin\apex{-1}~(16 atoms cell)\apex{-1}	& \AA\apex{-1}	&10\apex{5}~m/s	\\
    \bottomrule
  \end{tabular*}
  
\caption{Main bandstructure parameters at the Fermi level ($E_F=0$) for the field-effect-doped hydrogenated diamond (111) surface at a doping level of $n\ped{dop}=6\times10^{14}$~cm$^{-2}$: Total DOS $N_{\sigma}(0)$, band-resolved DOS $N_{\sigma,j}(0)$, band-averaged Fermi wavevector $\langle k\ped{F} \rangle$, band-resolved Fermi wavevector $\langle k\ped{F} \rangle_j$, band-averaged Fermi velocity $\langle v\ped{F} \rangle$, and band-resolved Fermi velocity $\langle v\ped{F} \rangle_j$. The indexes $j=1,2,3$ are labelled in Fig.~\ref{fig:el_phon_DFT}(b).} \label{tab:DOS}
\end{table}

We then performed linear response computations on a coarse grid of phonon momenta ($\vb{q}$-points) in the Brillouin zone in order to compute the electron-phonon (e-ph) matrix elements $g_{j\vb{k},j'\vb{k}'}^\nu$ for the phonon mode $\nu$ and between electronic states $\ket{j\vb{k}}$ and $\ket{j'\vb{k}'}$:
\begin{equation}
\label{eq:ep_matrix}
g_{j\vb{k},j'\vb{k}'}^{\nu} =\sum_{A\alpha} \frac{e^{A\alpha}_{\vb{k}'-\vb{k}\nu}}{\sqrt{2M_A\omega_{\vb{k}'-\vb{k}\nu}}}\bra{\vb{k}j}\fdv{v_{\text{SCF}}}{u_{A\alpha}^{\vb{k}'-\vb{k}}}\ket{j'\vb{k}'}
\end{equation}
where $j$ is the band index, $\vb{k}'=\vb{k}+\vb{q}$, and $v_{\text{SCF}}=V_{\text{KS}}e^{-i(\vb{k}'-\vb{k})\cdot\vb{r}}$ is the periodic part of the Kohn-Sham potential. $A$ labels atoms in the unit cell, $M_A$ is the mass of the $A$-th atom and $\alpha=\text{x,y,z}$ (i.e. the cartesian coordinates). $u_{A\alpha}^{\vb{k}-\vb{k}'}$ is the Fourier-transformed displacement of the $A$-th atom along $\alpha$ and the cartesian component of the normalized phonon eigenvector on the unit cell is $e^{A\alpha}_{\vb{k}'-\vb{k}\nu}$. Then, we exploited Wannier functions~\cite{MostofiCPC2014} in order to interpolate the e-ph matrix elements over the whole Brillouin zone, as described in Ref.~\onlinecite{CalandraPRB2010}. By doing so it was possible to compute the band-resolved Eliashberg spectral function $\alpha^2F_{jj'}(\omega)$ [Fig.~\ref{fig:el_phon_DFT}(c)] which gives the frequency-dependent e-ph interaction~\cite{SannaPRB2012}:
\begin{equation}
\label{eq:eliashberg_function}
\begin{split}
\alpha^2F_{jj'}(\omega)&=\frac{1}{N^2_{j,\sigma}(0)N_kN_q}\sum_{\vb{k}\vb{k}'}\alpha^2F_{j\vb{k},j'\vb{k}'}(\omega)\delta(\epsilon_{j\vb{k}})\delta(\epsilon_{j'\vb{k}'})\\
&=\frac{1}{N_{j,\sigma}(0)N_kN_q}\sum_{\vb{k}'\nu}\delta(\omega-\omega_{\vb{q}\nu})~\cdot\\
&\quad\cdot~\sum_{\vb{k}}\abs{g_{j\vb{k}j'\vb{k}'}^{\nu}}^2\delta(\epsilon_{j\vb{k}})\delta(\epsilon_{j'\vb{k}'})
\end{split}
\end{equation}
where $N_k$ ($N_q$) is the total number of $\vb{k}$-points ($\vb{q}$-points) used in the sum. The Dirac-delta $\delta(\epsilon_{j\vb{k}})$ and $\delta(\epsilon_{j'\vb{k}'})$ limit the scattering events to the Fermi surface. We also calculate the band-resolved e-ph coupling constant $\lambda_{jj'}$, i.e. the average strength of the e-ph interaction involving scattering events between the $j$-th and $j'$-th bands:
\begin{equation}
\label{eq:lambda}
\lambda_{jj'} = 2\int d\omega \frac{\alpha^2F_{jj'}(\omega)}{\omega}
\end{equation}
such that the total e-ph coupling constant, $\lambda$, is found as the weighted average of the band-resolved e-ph couplings, $\lambda_{jj'}$, by the corresponding band-resolved DOS at the Fermi level, $N_{\sigma,j}(0)$:
\begin{equation}
\label{eq:lambda_tot}
\lambda=\frac{\sum_{jj'}N_{\sigma,j}(0)\lambda_{jj'}}{\sum_{j}N_{\sigma,j}(0)}
\end{equation}
In our case we found that the total e-ph coupling constant is $\lambda=0.81$, while its decomposition on the electronic bands ($j,j'=1,2,3$ for the present system) gives a $3\times3$ matrix:
\begin{equation}
\label{eq:lambda_matrix}
\lambda_{jj'}=\left(\begin{array}{ccc}
                        0.4629 & 0.2425 & 0.1984\\
                        0.3973 & 0.2167 & 0.1806\\
                        0.3340 & 0.1856 & 0.1484\\
                      \end{array}
                    \right)
\end{equation}
The SC critical temperature $T_{\text{c}}$ was estimated using the McMillan/Allen-Dynes~\cite{McMillanPR1968,AllenPRB1975} formula:
\begin{equation}
\label{eq:AD}
T_{\text{c}} = \frac{\omega_{\textrm{log}}}{1.2}\exp{-\frac{1.04(1+\lambda)}{\lambda - \mu^*(1+0.62\lambda)}}
\end{equation}
where $\omega_{\textrm{log}}$ is the logarithmic-averaged phonon frequency, computed to be $\omega_{\textrm{log}}=83$~meV, and $\mu^*$ is the Morel-Anderson pseudopotential~\cite{MorelPR1962}, i.e. a measure of the effective e-e interaction. Since $\mu^*$ is an ad-hoc parameter, we assumed that its value falls within the same range as that of SC boron-doped bulk diamond~\cite{BlasePRL2004}, that is $\mu^{*}\in[0.13;0.14]$. The field-doped hydrogenated (111) diamond surface is then predicted to undergo a SC phase transition in the range of $T\ped{c}\in[29.6;34.9]$~K. In order to highlight the difference between inter- and intra-band couplings, we consider the symmetrized electron-phonon coupling constant matrix $\tilde{\lambda}_{jj'}=N_{\sigma,j}(0)\lambda_{jj'}/N_\sigma(0)$:
\begin{equation}
\label{eq:lambda_matrix_symm}
\lambda_{jj'}=\left(\begin{array}{ccc}
                        0.21 & 0.11 & 0.09\\
                        0.11 & 0.06 & 0.05\\
                        0.09 & 0.05 & 0.04\\
                      \end{array}
                    \right)
\end{equation}
where the diagonal elements describe the intra-band couplings, while the off-diagonal elements describe the inter-band ones. From Eq.~\ref{eq:lambda_matrix_symm} we can see that the most relevant processes are those concerning the intra-band scattering on the first Fermi surface (FS1$\leftrightarrow$FS1) and the inter-band scattering of the second (FS2$\leftrightarrow$FS1) and third Fermi surfaces (FS3$\leftrightarrow$FS1) with the first one.

\subsection{Imaginary-axis Migdal-Eliashberg equations}

The multi-band fully-anisotropic $\vb{k}$-resolved Eliashberg equations on the imaginary axis~\cite{EliashbergSovPhys1960, EliashbergZhTeor1960, AllenMitrovicSolStat1982, CarbotteRevModPhys, Ummarinorev, UmmarinoPhysicaC} consist in a set of $3j$ non-linear coupled equations which have to be solved self-consistently:
\begin{subequations}
\label{eq:Eli_equations}
\begin{eqnarray}
Z_{j\vb{k}}(i\omega_{n}) &&= 1+\frac{1}{\beta N_\sigma(0)\omega_{n}}\sum_{\vb{k}'n'j'}\frac{\omega_{n'}Z_{j'\vb{k}'}(i\omega_{n'})}{\Xi_{j'\vb{k}'}^2(i\omega_{n'})}\cdot\nonumber\\
&&~\cdot\lambda_{j\vb{k},j'\vb{k}'}(\omega_{n}-\omega_{n'})\label{eq:Eli_z}\\
\nonumber\\
\phi_{j\vb{k}}(i\omega_{n}) &&= \frac{1}{\beta N_\sigma(0)}\sum_{\vb{k}'n'j'}\frac{\phi_{j'\vb{k}'}(i\omega_{n'})}{\Xi_{j'\vb{k}'}^2(i\omega_{n'})}\cdot\nonumber\\
&&~\cdot[\lambda_{j\vb{k},j'\vb{k}'}(\omega_{n}-\omega_{n'})-\mu_{jj'}^*\theta(\omega-\abs{\omega_c})]\label{eq:Eli_phi}\\
\nonumber\\
\chi_{j\vb{k}}(i\omega_{n}) &&= -\frac{1}{\beta N_\sigma(0)}\sum_{\vb{k}'n'j'}\frac{\epsilon_{j'\vb{k}'}+\chi_{j'\vb{k}'}(i\omega_{n'})}{\Xi_{j'\vb{k}'}^2(i\omega_{n'})}\cdot\nonumber\\
&&~\cdot\lambda_{j\vb{k},j'\vb{k}'}(\omega_{n}-\omega_{n'})\label{eq:Eli_chi}
\end{eqnarray}
\end{subequations}
where $\omega_{n}=(2n+1)/\beta$ are the Matsubara's frequencies for fermions (with $n$ an integer number), $\beta=1/T$ is the inverse temperature (we set the Boltzmann constant $k_B=1$), $\mu_{jj'}^*$ is the effective Coulomb interaction between electrons on the $j$-th band and those on the $j'$-th band and the energies are measured from the Fermi level ($E_F=0$). $\Xi_{j}^2(\vb{k},i\omega_{n})$ is defined as:
\begin{equation}
\label{eq:chi}
\begin{split}
\Xi_{j\vb{k}}^2(i\omega_{n})=&~\omega_n^2Z_{j\vb{k}}^2(i\omega_{n}) +\\
 +&(\epsilon_{j\vb{k}}+\chi_{j\vb{k}}(i\omega_{n}))^2 + \phi_{j\vb{k}}^2(i\omega_{n})
\end{split}
\end{equation}
In order to avoid the divergence of the sums over $n$ in Eq.~\ref{eq:Eli_equations}, the number of Matsubara's frequency is limited by a cut-off energy $\omega_c$ which goes typically from $\omega_c\sim4\omega_{\text{max}}$ to $\omega_c\sim10\omega_{\text{max}}$  (with $\omega_{\text{max}}$ the maximum phonon frequency). In our case we choose $\omega_c=1290$ meV.
The e-ph coupling constant is then given by:
\begin{equation}
\label{eq:lambda_eli}
\begin{split}
\lambda_{j\vb{k},j'\vb{k}'}(\omega_{n}-\omega_{n'})=&\int_{0}^{\omega_{\text{max}}}d\omega\frac{2\omega}{(\omega_n-\omega_{n'})^2+\omega^2}\cdot\\
&\cdot\alpha^2F_{j\vb{k},j'\vb{k}'}(\omega)
\end{split}
\end{equation}
Moreover, in order to preserve the number of electrons in the system, the Fermi level must be re-computed self-consistently according to:
\begin{equation}
\label{eq:number}
N = 1-\frac{2}{\beta}\sum_{\vb{k}nj}\frac{\epsilon_{j\vb{k}}+\chi_{j\vb{k}}(i\omega_{n})}{\Xi_{j\vb{k}}^2(i\omega_{n})}
\end{equation}
Therefore, in the present work we would need to solve self-consistently a set of $10$ equations for the fully anisotropic system. Finally, the order parameter (i.e. the SC energy gap) for each band is defined as:
\begin{equation}
\label{eq:gap}
\Delta_{j\vb{k}}(i\omega_{n})=\frac{\phi_{j\vb{k}}(i\omega_{n})}{Z_{j\vb{k}}(i\omega_{n})}
\end{equation}
The order parameters have to be computed for different values of the temperature $T$, i.e. the set of equations Eq.~\ref{eq:Eli_equations} have to be solved for each value of $T$. The SC critical temperature $T\ped{c}$ is found when all of the $j$ energy gaps are zero, i.e. when $\Delta_{j\vb{k}}(i\omega_n;T_{\text{c}})=0$.\\

\begin{figure}[]
\centering
     \includegraphics[width=\mysmallwidth]{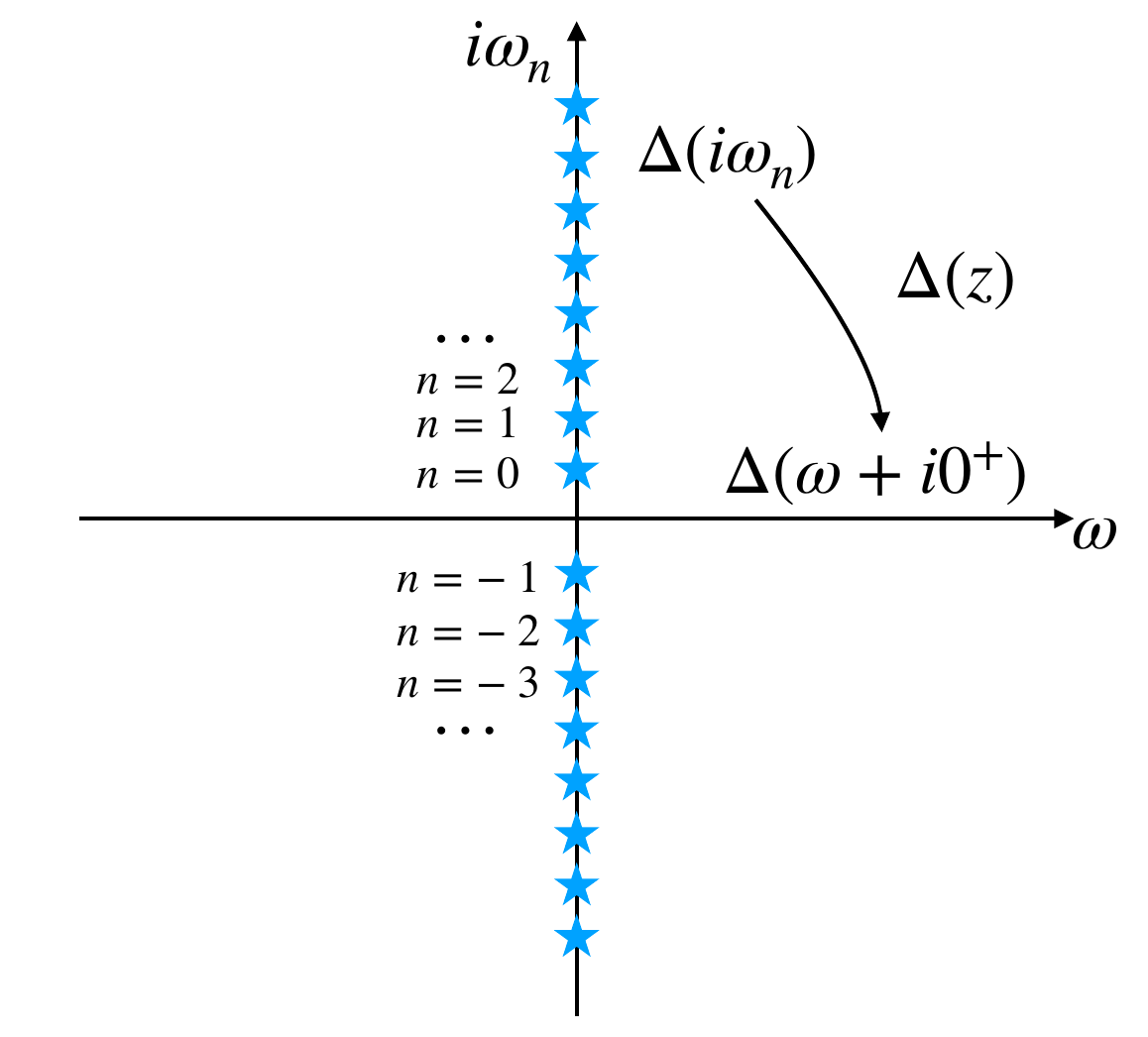}
      \caption
      {Analytic continuation of the superconducting energy gap from the imaginary axis ($\Delta(i\omega_{n})$) to the real axis ($\Delta(\omega +i0^+)$). Blue stars denotes the values of $\Delta(i\omega_{n})$ for $n=\dots,-2,-1,0,1,2,\dots$.}
   \label{fig:anal_cont}
\end{figure}

In order to simplify the computational cost, the DOS is assumed to be constant in an energy region of $2\hbar\omega_{\text{max}}$ around the Fermi energy. Therefore we can restrict ourselves only to a small energy range around $E\ped{F}$: in such a way Eq.~\ref{eq:Eli_chi} is exactly zero and Eq.~\ref{eq:number} can be neglected, reducing the problem to the self-consistent solution of only $6$ non-linear coupled equations. Moreover, for smooth and well-behaved (i.e non-crossing and isotropic) Fermi surfaces, we can pass from a $\vb{k}$-resolved picture to an energy description by averaging Eq.~\ref{eq:Eli_z} and Eq.~\ref{eq:Eli_phi} over the the Fermi surface. In this case we are left to solve the isotropic ME equations:
\begin{subequations}
\label{eq:Eli_equations_iso}
\begin{eqnarray}
&&Z_j(i\omega_n)=1+\frac{\pi}{\beta\omega_n}\sum_{n'j'}\frac{\omega_{n'}}{S_{j'}(i\omega_{n'})}\cdot\nonumber\\
&&\cdot\lambda_{jj'}(i\omega_n,i\omega_{n'})\\
\nonumber\\
&&Z_j(i\omega_n)\Delta_j(i\omega_n)=\frac{\pi}{\beta}\sum_{n'j'}\frac{\Delta_{j'}(i\omega_{n'})}{S_{j'}(i\omega_{n'})}\cdot\nonumber\\
&&\cdot\Bigl[\lambda_{jj'}(i\omega_n,i\omega_{n'})-\mu_{jj'}^*\theta(\omega-\omega_c)\Bigr]
\end{eqnarray}
\end{subequations}
Thanks to Pad\'e approximants~\cite{BakerPade1975} it is possible to analitically continue~\cite{VidebergJLTP1977, LeavensSSC1985} the SC gap (Eq.~\ref{eq:gap}) from the imaginary axis to the real axis (Fig.~\ref{fig:anal_cont}).
Once we have the frequency dependence of the order parameter along the real axis, it is possible to extract the quasiparticle DOS in the SC phase $N_{S,\sigma j}(\omega)$ through:
\begin{equation}
\label{eq:dos_qp}
\frac{N_{S,\sigma j}(\omega)}{N_{\sigma j}(0)}=\mathfrak{Re}\left[\frac{\omega}{\sqrt{\omega^2-\Delta_{j}^2(\omega)}}\right]
\end{equation}
Notice that in the single-band case the label $j$ is no longer needed and we simply have $\mu^*_{jj'}=\mu^*$.

\subsection{Choice of the Coulomb pseudopotential $\mu_{jj'}^*$}

The only free parameter in the ME theory is the effective Coulomb e-e repulsion, which is different~\cite{CarbotteRevModPhys} from the Morel-Anderson pseudopotential of the McMillan/Allen-Dynes formula. Similarly to the e-ph coupling, the effective Coulomb pseudopotential is a scalar in the single-band model and a matrix in the multi-band model, in principle resulting in a large number of free parameters. In order to keep the problem tractable, in the latter case we must make an Ansatz on the form of the band-resolved $\mu^{*}_{jj'}$. Its decomposition on the electronic bands is the following $3\times 3$ matrix:
\begin{equation*}
\left(\begin{array}{ccc}
                        \mu^{*}_{11} & \mu^{*}_{12}=c\cdot \mu^{*}_{11} & \mu^{*}_{13}=c\cdot \mu^{*}_{11}\\
                        \mu^{*}_{21}=\frac{N_{\sigma,1}(0)}{N_{\sigma,2}(0)}\mu^{*}_{12} & \mu^{*}_{22}=\mu^{*}_{11} & \mu^{*}_{23}=c\cdot \mu^{*}_{11}\\
                        \mu^{*}_{31}=\frac{N_{\sigma,1}(0)}{N_{\sigma,3}(0)}\mu^{*}_{13} & \mu^{*}_{32}=\frac{N_{\sigma,2}(0)}{N_{\sigma,3}(0)}\mu^{*}_{23} & \mu^{*}_{33}=\mu^{*}_{11}\\
                      \end{array}
                    \right)\\
\label{eq:multi_mu}
\end{equation*}
where $c$ is a number that we choose equal to either $0$, $\text{1/2}$ or $1$. For example, in the case of MgB$_{2}$ (the most known two-band e-ph superconductor) this parameter is approximately $0<c<\text{1/2}$~\cite{UmmarinoPhysC2004}. We can then retrieve a total Coulomb pseudopotential $\mu_{\text{tot}}^*$ as:
\begin{equation}
\mu_{\text{tot}}^{*}=\frac{\sum_{jj'}N_{\sigma,j}(0) \mu^{*}_{jj'}}{\sum_{j}N_{\sigma,j}(0)}
\label{eq:mu_tot}
\end{equation}
We estimate the value of $\mu^*\ped{tot}$ most appropriate for the ME formalism by first solving the single-band isotropic ME equations (Eq.~\ref{eq:Eli_equations_iso}) in the case of boron-doped bulk diamond, which is known to undergo a SC phase transition at $T\ped{c}\approx4$ K at a boron concentration of \mbox{$\sim1.85\%$}. Using the Eliashberg spectral function computed in Ref.~\onlinecite{BlasePRL2004}, the experimental value of $T\ped{c}$ can be found by setting $\mu^*\ped{tot}=\mu^*=0.17$. This value will be considered representative also for the hydrogenated diamond (111) surface. In the multi-band case, the values of $\mu^*_{jj'}$ are determined by $\mu^*\ped{tot} = 0.17$ via Eq.~\ref{eq:mu_tot} and the chosen value of $c$, as summarized in Table~\ref{tab:mu_star}.
\begin{table}[h]
  \begin{tabular*}{\mywidth}{@{\extracolsep{\fill}}cccccc}
  \toprule
    & Single-band & \multicolumn{3}{c}{Multi-band} \\
    \midrule
     &  & $c=0$ & $c=1/2$ & $c=1$\\
    \midrule
     $\mu^*\ped{tot}$ & 0.17 & 0.17 & 0.17 & 0.17 \\
     $\mu^*_{jj',\,j=j'}$ & /  & 0.17 & 0.07785 & 0.05048  \\
     $\mu^*_{jj',\,j<j'}$ & /  & 0 & 0.038925 & 0.05048  \\
    \bottomrule
  \end{tabular*}
\caption{Coulomb pseudopotentials employed in the solution of the single-band and multi-band Migdal-Eliashberg equations. All values are obtained under the condition that the total pseudopotential reproduces the experimental $T\ped{c}$ of boron-doped bulk diamond. The indexes $j=1,2,3$ are labelled in Fig.~\ref{fig:el_phon_DFT}(b).} \label{tab:mu_star}
\end{table}
%

\section{\label{sec:results}Results}

\subsection{Isotropic single-band solution}
\begin{figure}[b]
\centering
     \includegraphics[width=\mywidth]{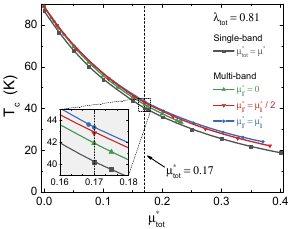}
      \caption
      {Superconducting critical temperature $T\ped{c}$ vs. the total effective Coulomb electron-electron interaction $\mu_{\text{tot}}^*$. The black line corresponds to the solution of the isotropic single-band ($j{=}j'{=}1$ and $\mu^*{=}\mu^*_{\text{tot}}$) ME equations (Eq.~\ref{eq:Eli_equations_iso}). The green, red and blue lines correspond to the solution of the isotropic multi-band ($j,j'=1,2,3$) Eliashberg equations with $c=0,1/2,1$ respectively for $\mu^*_{jj'}$.}
   \label{fig:mu_star}
\end{figure}
\begin{figure}
\centering
     \includegraphics[width=\mysmallwidth]{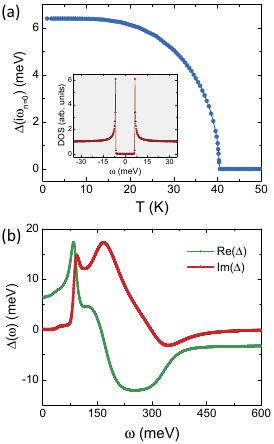}
      \caption
      {(a) Superconducting order parameter on the imaginary-axis, $\Delta(i\omega_{n=0})$, as a function of temperature $T$ obtained from the solution of the isotropic single-band Migdal-Eliashberg equations with $\mu^*=0.17$. The inset shows the single-band quasiparticle density of states (DOS) as a function of energy $\omega$; (b) Real (Re) and Imaginary (Im) part of the superconducting order parameter on the real-axis, $\Delta(\omega)$, obtained from the analytic continuation of $\Delta(i\omega_{n})$ computed at $T=T_{c}/10$.}
   \label{fig:single_band}
\end{figure}

We start our analysis from a single-band isotropic description of our system, and compute the resulting dependence $T\ped{c}$ for increasing values of $\mu^*$ (solid black line in Fig.~\ref{fig:mu_star}). For a given value of $\mu^*$, this is done by computing the $T$-dependence of the SC order parameter on the imaginary axis $\Delta(i\omega_{n=0})$, as shown in Fig.~\ref{fig:single_band}(a), and $T\ped{c}$ is then found as the value of $T$ for which $\Delta(i\omega_{n})\rightarrow0$
. For the representative value $\mu^*=0.17$, we find $T\ped{c}=40.3$~K. Note that this value is higher than the one obtained from the McMillan/Allen-Dynes formula ($T\ped{c}{\in}[29.6;34.9]$, $\mu^*{\in}[0.13;0.14]$), indicating that the latter significantly underestimates the value of $T\ped{c}$ in our system. We determine the value of the SC order parameter in the $T\rightarrow0$ limit, obtaining $\Delta(i\omega_{n=0};T{=}0)=6.46$~meV, from which we can calculate the gap-to-$T\ped{c}$ ratio:
\begin{equation}
\label{eq:single_ratio}
r=\frac{2\Delta(i\omega_{n=0};T=0)}{k_BT_{c}}=3.72
\end{equation}
which is close to the BCS value of $r_{\text{BCS}}=3.54$. Usually the gap-ratio (Eq.~\ref{eq:single_ratio}) requires the value of the superconducting gap computed on the real axis, however in our case the one computed on the imaginary-axis is not different (i.e. we are in a weak-coupling limit). We then perform the analytic continuation of the SC gap on the imaginary-axis at $T=T\ped{c}/10$, $\Delta(i\omega_{n};T=T\ped{c}/10)$, to the real-axis [$\Delta(\omega)$, Fig.~\ref{fig:single_band}(b)]. From this we finally compute the quasiparticle DOS [see inset to Fig.~\ref{fig:single_band}(a)], which shows the typical shape of an $s$-wave superconductor.

\subsection{Isotropic multi-band solution}

\begin{figure*}[]
\centering
     \includegraphics[width=\textwidth]{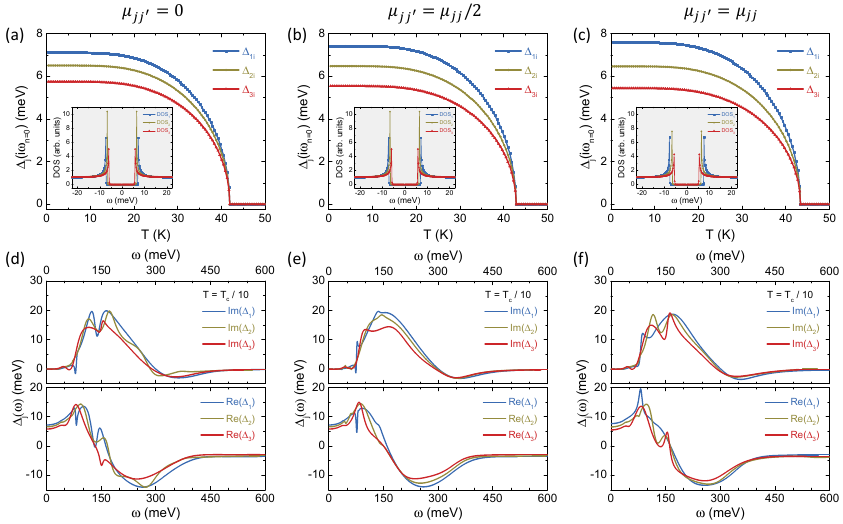}
      \caption
      {(a-c) Band-resolved superconducting order parameters $\Delta_{j}(i\omega_{n=0})$ [$j=1,2,3$ as labelled in Fig.~\ref{fig:el_phon_DFT}(b)] on the imaginary-axis as a function of temperature $T$ obtained from the solution of the isotropic multi-band Migdal-Eliashberg equations at $\mu^*\ped{tot}=0.17$ using (a) $c=0$, (b) $c=1/2$, and (c) $c=1$ for $\mu^*_{jj'}$. In the inset we show the multi-band quasiparticle densities of states DOS$_j$ as a function of energy $\omega$ calculated at $T=T\ped{c}/10$; (d-f) Real (Re) and Imaginary (Im) part of the superconducting order parameters $\Delta_j(\omega)$ on the real-axis obtained from the analytic continuation of $\Delta_j(i\omega_{n})$ computed at $T=T\ped{c}/10$ using $\mu^*\ped{tot}=0.17$ and (d) $c=0$, (e) $c=1/2$, and (f) $c=1$ for $\mu^*_{jj'}$.}
   \label{fig:multi_iso}
\end{figure*}

Having firmly established the description of our system in the single-band approximation, we now move to the more accurate multi-band description. In this case, the behavior of the system depends on the structure of the band-resolved Coulomb pseudopotential $\mu_{jj'}^*$, which in turn is given by the chosen value of $c$. The simplest choice we can make is $c=0$, i.e. neglecting any repulsion between charge carriers pertaining to different bands (inter-band repulsion). We plot the corresponding dependence of $T\ped{c}$ on $\mu^*\ped{tot}$ as a solid green line in Fig.~\ref{fig:mu_star}, and we find that the values of $T_{c}$ are slightly higher than those calculated in the single-band approximation. We also note that the solution of Eq.~\ref{eq:Eli_equations_iso} becomes unstable for an equivalent $\mu^*_{\text{tot}}\gtrsim0.23$, much larger than the physically-representative $\mu^*\ped{tot}=0.17$ for which we find $T\ped{c}=41.9$~K. Indeed, the most stable solution to the multi-band ME equations is found by setting $c=1/2$, as it can be extended up to $\mu^*\ped{tot}\approx0.38$.~We plot the resulting dependence of $T\ped{c}$ on $\mu^*\ped{tot}$ as the solid red line in Fig.~\ref{fig:mu_star}, from which we find a further increase in $T\ped{c}$ with respect to both the single-band approximation and the multi-band $c=0$ limit. For this intermediate strength of the inter-band repulsion, at the representative $\mu^*\ped{tot}=0.17$ we find $T\ped{c}=42.9$~K. Finally, we consider the opposite limit $c = 1$, where inter- and intra-band repulsions have exactly the same strength, and plot the corresponding $\mu^*\ped{tot}$-dependence of $T\ped{c}$ as as the solid blue line in Fig.~\ref{fig:mu_star}. For $\mu^*\ped{tot}\lesssim 0.12$, this solution is basically equivalent to that at $c=1/2$. At larger values of $\mu^*\ped{tot}$, $T\ped{c}$ is found to slightly increase over the $T\ped{c}$ calculated at $c = 1/2$, although the solution to Eq.~\ref{eq:Eli_equations_iso} becomes unstable at a slightly lower $\mu^*\ped{tot}\approx0.37$. In this strong inter-band repulsion limit, we find $T\ped{c} = 43.4$~K at the representative $\mu^*\ped{tot}=0.17$. For all considered choices of $c$, the $T\ped{c}$ calculated using the multi-band ME equations is larger than the one calculated using the single-band model: This indicates that the multi-band pairing strengthens the SC phase in the system, a behavior observed in several different superconductors~\cite{UmmarinoPhysC2004, PickettBook, GonnelliSciRep2016, PiattiNL2018}. In the case of field-effect-doped (111) diamond, our calculations show that the multi-band pairing enhances $T\ped{c}$ between $\approx4-8\%$ at $\mu^*\ped{tot} = 0.17$ with respect to single-band paring, depending on the relative weight between intra-band and inter-band repulsions. More specifically, shifting the repulsion weight from intra-band terms ($c=0$ limit) to inter-band terms ($c=1$ limit) is beneficial to $T\ped{c}$, consistently with the largest element in the band-resolved e-ph coupling $\lambda_{jj'}$ being the intra-band element $\lambda_{11}$ (which couples charge carriers within FS1).
\begin{table}[]
  \begin{tabular*}{\mywidth}{@{\extracolsep{\fill}}llllll}
  \toprule
    & & Single-band & \multicolumn{3}{c}{Multi-band} \\
    \midrule
     & &  & $c=0$ & $c=1/2$ & $c=1$ \\
    \midrule
     $T\ped{c}$ & K & 40.3 & 41.9 & 42.9 & 43.4 \\
     $\Delta$ & meV & 6.46  & / & / & / \\
     $\Delta_{1}$ & meV & /  & 7.11 & 7.43 & 7.59 \\
     $\Delta_{2}$ & meV & /  & 6.50 & 6.47 & 6.46 \\
     $\Delta_{3}$ & meV & /  & 5.74 & 5.54 & 5.44 \\
     $r$ & & 3.72 & 3.62 & 3.60 & 3.58 \\
     $r_1$ & & / & 3.94 & 4.02 & 4.06 \\
     $r_2$ & & / & 3.60 & 3.50 & 3.45 \\
     $r_3$ & & / & 3.12 & 3.00 & 2.91 \\
    \bottomrule
  \end{tabular*}
\caption{Superconducting (SC) critical temperatures $T\ped{c}$, SC order parameters for $T{\rightarrow0}$ on the imaginary-axis $\Delta_{j}(i\omega_{n=0};T{=}0)$, and gap-to-$T\ped{c}$ ratios $r_j$ (Eq.~\ref{eq:ratios}), as obtained from the solution of the isotropic ME equations with $\mu^*\ped{tot}{=}0.17$. The multi-band cases are calculated using $c=0$,$c=1/2$ and $c=1$ for $\mu^*_{jj'}$. The indexes $j=1,2,3$ are labelled in Fig.~\ref{fig:el_phon_DFT}(b).} \label{tab:iso_summary}
\end{table}

We now consider the effect of multi-band pairing on the SC order parameter. In Fig.~\ref{fig:multi_iso}(a-c) we plot the three SC gaps $\Delta_{j}(i\omega_{n})$ [$j=1,2,3$, as labelled in Fig.~\ref{fig:el_phon_DFT}(b)] computed by solving the isotropic multi-band ME equations along the imaginary-axis with $\mu^*\ped{tot}=0.17$ and $c=0$ (a), $c=1/2$ (b), and $c=1$ (c). The presence of multi-band pairing not only increases the $T\ped{c}$, as we discussed previously, but also lifts the degeneracy between the three SC gaps as $T\rightarrow0$. For any value of $c$, we find that $\Delta_{1}(i\omega_{n=0};T{=}0)$ is larger than the single-band $\Delta(i\omega_{n=0};T{=}0)$, and that $\Delta_{3}(i\omega_{n=0};T{=}0)$ is smaller. For $c=0$, $\Delta_{2}(i\omega_{n=0};T{=}0)$ is slighly larger than $\Delta(i\omega_{n=0};T{=}0)$, whereas for both $c=1/2$ and $c=1$ they are almost equal to each other. Overall, the splitting between the SC gaps increases as the weight of inter-band repulsion increases. For each SC gap, we also compute the corresponding gap-to-$T\ped{c}$ ratio:
\begin{equation}
\label{eq:ratios}
r_j=\frac{2\Delta_j(i\omega_{n=0};T=0)}{k_BT_{c}}
\end{equation}
from which we determine that in second band $r_2$ is close to the BCS value $r\ped{BCS} = 3.54$ for any value of $c$, whereas in the first and third bands $r_1$ and $r_3$ deviate almost symmetrically away from it, another typical behavior exhibited by multi-band superconductors~\cite{InosovPRB2011}. Namely, $r_1>r\ped{BCS}$ and increases at the increase of the inter-band repulsion weight, whereas $r_3<r\ped{BCS}$ and decreases at the increase of the inter-band repulsion weight. The results of the multi-band model are consistent with those found single-band picture, as can be assessed by computing the weighted average of $r_j$ over the DOS per band $N_{\sigma,j}$:
\begin{equation}
r = \frac{\sum_{j}N_{\sigma,j}(0) r_j}{\sum_{j}N_{\sigma,j}(0)}
\end{equation}
We obtain $r=3.62$ for $c=0$, $r=3.60$ for $c=1/2$, and $r=3.58$ for $c=1$, close (even if smaller) to the single-band ratio $r=3.72$ we computed via Eq.~\ref{eq:single_ratio}, and decreasing at the increase of the inter-band repulsion weight. We summarize all the main parameters obtained by solving both the single- and multi-band isotropic ME equations in Table~\ref{tab:iso_summary}.

In Fig.~\ref{fig:multi_iso}(d-f) we plot the analytic continuations on the real axis $\Delta_j(\omega)$ [$j=1,2,3$ as labelled in Fig.~\ref{fig:el_phon_DFT}(b)] of the SC order parameter computed on the imaginary axis at $T=T_{c}/10$ and for the three different choices of inter-band repulsion weight: $c=0$ (d), $c=1/2$ (e), and \mbox{$c=1$ (f)}. From these, we obtain the quasiparticle DOS for each band, which are plotted in the insets to \mbox{Fig.~\ref{fig:multi_iso}(a-c)}. 
{\color{blue}Note that our choice to focus on $T=T_{c}/10$ is not arbitrary: It is based on $T=T_{c}/10$ being the standard condition for the reliable detection of multi-band features in the experimental tunnelling and Andreev-reflection spectroscopy data~\cite{DagheroSUST2010, GonnelliPRL2002, GonnelliPRL2008, DagheroSUST2018}, and on the fact that computing the quasiparticle DOS at higher $T$ is beyond the validity of the analytic continuation and would thus require the solution of the much more computationally demanding ME equations on the real axis~\cite{VidebergJLTP1977, LeavensSSC1985}.}
Once again, the presence of multi-band pairing is observed as a splitting of the single-peaked quasiparticle DOS computed in the single-band approximation into three different peaks, one for each SC gap. For all values of $c$, the quasiparticle DOS maintain the typical U-shape of s-wave superconductors, with larger inter-band repulsion weights leading to a more pronounced splitting between the peaks. From the experimental point of view, the detection of a multi-peaked quasiparticle spectrum would be a direct evidence for the multi-band nature of the pairing, and would allow to determine the actual value of the inter-band repulsion weight. This could be achieved by performing tunnelling spectroscopy measurements between buried boron-doped diamond electrodes electrically separated from the field-induced channel at the surface, similarly to the work described in Ref.~\onlinecite{CostanzoNatNano2018}.

{\color{blue}Concerning the experimental feasibility of the proposed architecture, the most significant challenge lies in attaining the necessary $n_{dop} = 6\times 10^{14}$~cm\apex{-2} to trigger the high-Tc SC state, as the current record for ion-gated hydrogenated intrinsic diamond is nearly one order of magnitude lower~\cite{AkhgarNL2016} due to the small quantum capacitance of diamond~\cite{PiattiLTP2019, PiattiEPJ2019, PiattiApSuSc2020}. Hole co-doping approaches have shown promise in this regard, combining ionic gating of hydrogenated surfaces with B substitution, and have allowed to reach $n_{dop} \sim 2\times 10^{14}$~cm\apex{-2}~\cite{PiattiLTP2019, PiattiEPJ2019, PiattiApSuSc2020}. Consequently, further improvements will be needed, such as developing ionic media capable of withstanding larger gate voltages, optimizing the B doping process to maximize the surface free carrier density, and increasing the quantum capacitance of the diamond surface by introducing an ultrathin spacer layer between diamond and the ionic gate~\cite{PiattiApSuSc2020, ZhangNL2019}.}

\section{\label{sec:conclusion}Conclusions}

In this work we have analyzed, by means of the isotropic Migdal-Eliashberg theory, the high-T\ped{c} superconducting transition induced at the hydrogenated (111) surface of diamond when doped with holes via the electric field effect at a concentration of $n\ped{dop}=6\times10^{14}$~cm$^{-2}$. Starting from the band-resolved electron-phonon spectral functions computed ab initio in a precedent work~\cite{RomaninAPSUSC2019}, we first give a single-band description of the system and then we further extend the solution of the ME equations in order to account for the multi-band nature of the electronic dispersion relation at the Fermi level. The total effective electron-electron interaction $\mu^*_{tot}$ is fixed by taking B-doped bulk diamond as reference, which gives $\mu^*=0.17$ according to the solution of the Migdal-Elishberg equations at a B doping concentration of $\sim1.85\%$ (i.e. $T_{\text{c}}\approx4$~K). In the single-band approximation we have that $\mu^*=\mu^*_{tot}$, while in the multi-band model we consider three possible Ansatz for the effective-coulomb interaction matrix $\mu^*_{jj'}$, always requiring that $\mu^*_{tot}=\sum_{jj'}N_{\sigma,j}(0) \mu^{*}_{jj'}/\sum_{j}N_{\sigma,j}(0)$. The solution of the isotropic single-band ME equations shows that the SC phase transition is attained at $\sim40$ K with an s-wave gap value of $\sim6.5$~meV. However, when we take into account multi-band effects the T\ped{c} is enhanced by $~4-8~\%$, depending on the specific Ansatz for $\mu^*_{jj'}$. For every choice of $\mu^*_{jj'}$ we have analyzed, the band-resolved SC gaps $\Delta_j$ show an s-wave symmetry with $\Delta_{1}>\Delta_{2}>\Delta_{3}$. To offer a guideline for future experimental works aimed at determining the properties of the field-induced SC state, we also calculated a quantity sensitive to the multi-band pairing, namely the quasiparticle density of states, where we found a clear signature of multi-band effects. 
Finally, we would like to point out that  -- while we have so far discussed the most resonable choices for the inter- and intra-band repulsion weights -- more specific choices also allow to stabilize SC pairings not featuring the standard s-wave symmetry. For example, a choice of Coulomb pseudopotential such that $\mu^{*}_{j3}=0.15$ and $\mu^{*}_{jk}=0$ leads to the stabilization of an $s\pm$ symmetry for the SC order parameter while keeping the source of pairing as purely e-ph driven.
%
%
%
%
\begin{acknowledgments}
The authors acknowledge M. Calandra for fruitful scientific discussions. G.A.U. acknowledges support from the MEPhI Academic Excellence Project (Contract No. 02.a03.21.0005). E.P. acknowledges funding from the MIUR PRIN-2017 program (Grant No. 2017Z8TS5B -- ``Tuning and understanding Quantum phases in 2D materials -- Quantum2D"). Computational resources were provided by hpc@polito (http://hpc.polito.it) and by CINECA, through the 'ISCRA C' project 'HP10C8P1FI'.
\end{acknowledgments}

\end{document}